\documentclass[a4paper,prl,twocolumn,floatfix,english]{revtex4}
\usepackage{epsfig, graphicx,psfrag,amsmath,amssymb,float}
\usepackage[T1]{fontenc}
\usepackage[latin9]{inputenc}
\usepackage{amsmath}
\usepackage{amssymb}
\usepackage{color}
\usepackage[svgnames]{xcolor}
\usepackage{amscd}
\usepackage{bm}
\usepackage{psfrag}
\usepackage{babel}
\usepackage{bbding}
\usepackage{epsfig}
\usepackage{graphicx}
\usepackage{dcolumn}
\usepackage{bm}
\usepackage{ulem}
\usepackage{lipsum}
\usepackage{amssymb}
\usepackage{amsmath}
\usepackage{dsfont}
\usepackage{soul,color}
\newcommand{\be}{\begin{equation}}
\newcommand{\ee}{\end{equation}}
\newcommand{\bea}{\begin{eqnarray}}
\newcommand{\eea}{\end{eqnarray}}

\begin{document}
\title{The dynamic structure factor in impurity-doped spin chains}

\author{Annabelle Bohrdt$^{1,2}$, Kevin J\"agering$^{1}$, Sebastian Eggert$^1$, and Imke Schneider$^1$}

\affiliation{$^1$Physics Department and Research Center OPTIMAS, University of Kaiserslautern, D-67663 Kaiserslautern, Germany}
\affiliation{$^2$Department of Physics and Institute for Advanced Study,
Technical University of Munich, 85748 Garching, Germany}

\begin{abstract}

The effects of impurities in spin-1/2 Heisenberg chains are recently 
experiencing a renewed interest due to experimental realizations in solid state systems
and ultra-cold gases.  The impurities effectively cut the chains into finite segments with 
a discrete spectrum and characteristic correlations, which have a
distinct effect on the dynamic structure factor.  Using bosonization and the numerical 
Density Matrix Renormalization Group we provide detailed quantitative predictions for
the momentum and energy resolved structure factor in doped systems. Due to the impurities,
spectral weight is shifted away from the antiferromagnetic wave-vector $k=\pi$ into regions
which normally have no spectral weight in the thermodynamic limit.  The effect can 
be quantitatively described in terms of scaling functions, which are derived from a recurrence
relation based on bosonization.

\end{abstract}

\maketitle

Spin chains 
have been the center of attention as prototypical quantum many body systems
ever since the early days of quantum mechanics \cite{bethe}
and up to this day significant advances
are made, e.g.~in describing exact form factors \cite{caux2005,caux2005b,caux2011,caux2012,kitanine2012}, exact correlations \cite{dugave}, non-equilibrium states \cite{prozen,nonequil}, 
and dynamic correlations 
in the regime of a non-linear spectrum \cite{imambekov2009,glazman2011,pereira2006,
pereira2007,pereira2008,cheianov2008,karrasch2015,eliens2016,karimi2011}.
Recently, there has been renewed experimental interest in intentionally doped 
spin chain systems \cite{doped,doped2} with new results on the Knight 
shift \cite{utz,utz2}, magnetic ordering \cite{ordering},  and the 
dynamic structure factor \cite{average,average2,neutron,neutron2}.
Doped spin chains are known to acquire characteristic boundary 
correlation functions \cite{eggert1992}, which lead to 
impurity induced changes in the Knight shift \cite{laflorencie,eggert1995}, 
the susceptibility \cite{susc1,susc2,susc3},
the static structure factor \cite{eggert1995}, and 
the ordering temperature \cite{kojima97,eggert2002}.
However, surprisingly
a systematic analysis of the doping effects on the energy and momentum resolved 
dynamic structure factor is still missing so far.  
Previous research has taken into account the discrete spectrum 
of finite chains \cite{wessel,average,average2}, which leads to an 
exponential suppression at low energies \cite{average,average2}. 
The understanding of the momentum dependence is more involved, however, since
characteristic correlations near the impurities play an 
important role and lead to 
a strong redistribution of spectral weight 
to higher momenta outside the dispersion relation as shown in this paper.

\begin{figure}[t]
\includegraphics[width=1\columnwidth,keepaspectratio]{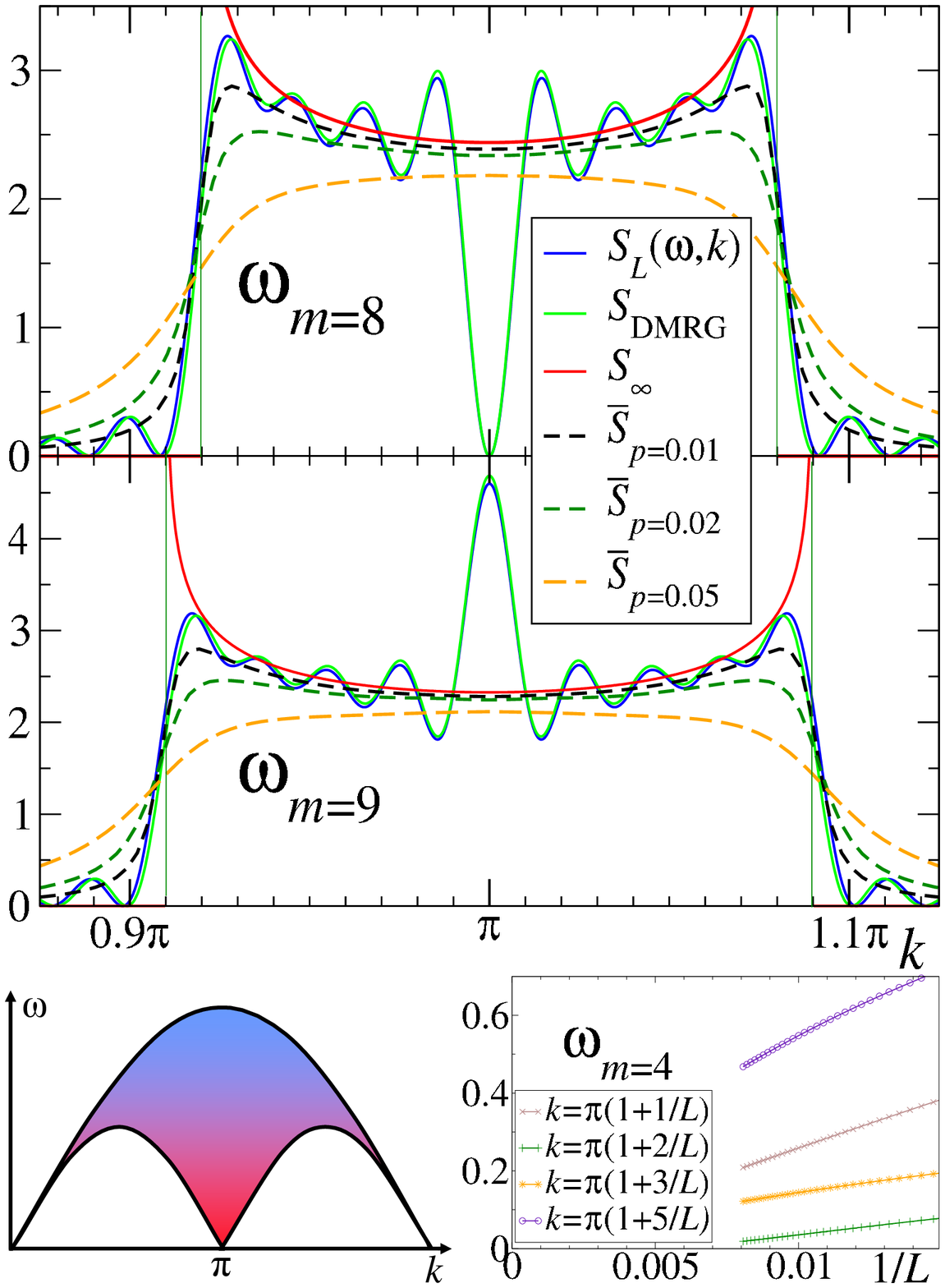}  
\caption{{\it Top:}
The dynamical structure factor $S_L(\omega_m,k)$ at $L=100$ as a function of $k$ near $\pi$
from bosonization compared to numerical DMRG calculations for 
$\omega_m=8\frac{\pi v}{L} \approx 0.31J$ and $9\frac{\pi v}{L} \approx 0.348J$.
The $L\to \infty$ behavior $S_\infty$ from Eq.~(\ref{bulk}) 
and the averaged signal $\bar S_p$ with doping level $p$
at the same energies are
also shown where vertical lines mark $v |k-\pi| = \omega_m$. 
{\it Bottom left:} Schematic spinon dispersion.
{\it Bottom right:} Bosonization error over finite size deviation $\Delta S$ in 
Eq.~(\ref{Delta}) for $m=4$ and different $k$ using $K=0.8$ as a function of $1/L$.
}\label{fig:level8and9}  
\end{figure}

The underlying model is the well-known
$xxz$-spin chain 
\begin{eqnarray} \label{model}
 H=J \sum_{i=1}^{L-1} \left( S_i^xS_{i+1}^x +  S_i^y S_{i+1}^y  + \Delta  S_i^z S_{i+1}^z \right)
\end{eqnarray}
which represents
a one-dimensional array of $L$ interacting spin-1/2 operators with open boundary conditions.
This model  can 
also be used to describe hard-core bosons \cite{hardcore}, quantum dimer systems \cite{dimer}, or triplon
excitations in ladder systems \cite{triplon}.
The longitudinal dynamic structure factor is a key quantity, which can be 
measured by angle-resolved neutron scattering experiments \cite{lake2005,lake2005b,lake2013}
and
at the same time gives a deep insight into the spatial-temporal correlations.
Impurities in the systems cut the spin chains \cite{eggert1992,wessel}, so we 
consider the structure factor for finite segments of
length $L$
\begin{eqnarray}
S(\omega,k)=\frac{1}{L}\sum_{j,j^\prime} e^{-i k (j-j^\prime)} \int_{-\infty}^\infty dt \,e^{i \omega t}\langle S^z_j(t)S^z_{j^\prime}(0)\rangle. \label{struct}
\end{eqnarray}
In recent years it was possible to calculate $S(\omega,k)$
to high accuracy
from exact methods in the thermodynamic limit \cite{caux2011,caux2012} which is nonzero 
only inside the bounds of the known dispersion  \cite{spinon}
as shown schematically in the bottom left panel 
of Fig.~\ref{fig:level8and9}.  Indicated in red are the
dominant correlations near
the antiferromagnetic wave-vector $k=\pi$ at low frequencies, which will be the topic 
of this paper.
The low-energy behavior for infinite chains $L\to \infty$
has been known since the 1980s 
from bosonization \cite{schulz}
and is described by powerlaws as also derived in the appendix   
\be \label{bulk}
S_\infty(\omega,q+\pi) = \left({2 v}\right)^{1-2K}\!\!{\pi^2 A_z}{ \Gamma^{-2}\!(K)}
\left(\omega^2-v^2q^2 \right)^{K-1}\!, 
\ee 
where $|q|=|k-\pi|<\omega/v$, 
$K=\pi/2(\pi-\theta)$ 
is the Luttinger parameter, 
$v=J \pi \sin\theta/2 \theta$ is the spinon velocity in terms of $\cos\theta=\Delta$ \cite{review}
and the overall amplitude $A_z$ is
known from exact methods \cite{lukyanov2002}.
Since $K <1$ for $\Delta > 0$ the signal increases with $\omega^{2K-2}$
as the frequency is lowered and shows a divergence near the dispersion 
$\omega^2-v^2 q^2\to 0^+$, but vanishes for $|vq|>\omega$.
A substantial amount of literature 
has been devoted to the analysis of the 
divergence at the dispersion in spin chains and 
quantum wires \cite{meden1992}, 
which finds that it is not universal but depends on 
non-linear effects \cite{imambekov2009,glazman2011,pereira2006,pereira2007,
pereira2008,cheianov2008,karrasch2015,eliens2016,karimi2011} as well as the cut-off procedure \cite{meden1999,markhof2016}.   
In this work we now consider finite chains, which show no divergence
at all.  Remarkably,  at low energies
bosonization nonetheless gives quantitatively accurate results
for all $q=k-\pi$ and $L$.
It is therefore possible to perform an efficient large-scale impurity averaging to
predict the experimental signal.

The low-energy spectrum of finite spin chains is 
well described by equally spaced
energy levels $\omega_m \approx m \Delta\omega$ with $\Delta \omega= \frac{\pi v}{L}$ \cite{eggert1992}.
Higher order corrections
to this quantization 
are well understood \cite{log1,sirker-bortz,bortz,eggert1992}, but do not 
change the averaged signal.
Because of the discrete spectrum it is useful
to rewrite 
Eq.~(\ref{struct}) in the Lehmann representation
\begin{eqnarray}
S(\omega,k)=
\Delta\omega \sum_{m\neq 0} S_L(\omega_m,k) \delta(\omega-\omega_m)
\label{lehmann}
\end{eqnarray}
where we have defined individual spectral weights
\be \label{SL}
S_L(\omega_m, k) =
\frac{2 \pi}{\Delta \omega} |\langle \omega_m | S_k^z|0 \rangle |^2
\ee
with $S_k^z=\frac{1}{\sqrt {L}}\sum_{j} e^{-i k j} S_j^z$.   
Note that the scattering wave-vectors $k$ are not quantized.

Numerically the spectral weights $S_L(\omega_m,k)$ can be 
evaluated by targeting a large number
of excited 
states in Density Matrix Renormalization Group (DMRG) simulations \cite{white}
as discussed below.
Analytically it turns out that bosonization gives an accurate estimate for $S_L$
from a simple sum over a finite number of collective modes, which agree
accurately with our DMRG simulations already for moderate lengths.

The bosonization and calculation of correlation functions of finite spin chains
has been discussed before \cite{eggert1992,mattsson,eggert2002,sirker-bortz,log} as reviewed in the appendix.  It is based on expressing the alternating part of the spin-operator
in terms of a free bosonic field $\phi$
\be S^z(x,t) \approx A (-1)^x \sin\sqrt{4 \pi K} \phi(x,t),\ee
where the amplitude $A^2= A_z/2$ is known from exact methods 
\cite{lukyanov2002}. 
Long-distance correlations can then be calculated by expectation values of the form
\be
G^{\pm}(x,y,t) = \langle e^{i\sqrt{4 \pi K}\phi(x,t)} e^{\mp i\sqrt{4 \pi K}\phi(y,0)}\rangle.
\ee
The main technical difficulty is
the Fourier-transform over time in Eq.~(\ref{struct}), which 
ordinarily requires a detailed analysis of the analytic structure and 
contour integrals with a cut-off procedure \cite{schulz,boundaryDOS,meden1992}.
However, in our calculation we use finite systems, which 
provides a efficient way of calculating spectral weights, that can 
be summarized in a few lines  as follows and is derived in the appendix.  
Due to the discrete energy spectrum, the Fourier-transform
gives a sum over delta-functions 
\be \int_{-\infty}^\infty dt G^\pm(x,y,t) e^{i\omega t} = 
2 \pi \sum_m S^\pm_m (x,y) \delta(\omega-\omega_m).  \label{int}
\ee
To evaluate the spectral weights $S_m^\pm$ it is possible to use
the mode expansion and an integration by parts of Eq.~(\ref{int})
to arrive at a recurrence relation \cite{schoenhammer1993,schneider2010}
\be\label{recursion}
S_{m}^\pm(x,y)  =  \frac{\pm 1}{m}\sum_{\ell=1}^m S_{m-\ell}^\pm(x,y) \,\gamma_\ell(x,y),
\ee
which allows to express the $S_m^\pm$ in Eq.~(\ref{int}) as 
a recursive sum of the ones with lower index $m-\ell$ using starting values of 
\be
S_0^+(x,y)\! =\!S_0^-(x,y)\! = \!c(x) c(y)\! =\!
\left(\frac{4L^2}{\pi^2}\sin\! \frac{\pi x}{L} \sin\!\frac{\pi y}{L}\right)^{-K}
\label{start}
\ee
and the coefficients
\be
\gamma_\ell(x,y)
= 
4 K \sin\frac{\ell \pi  x}{L} \sin\frac{\ell \pi  y}{L}.
\label{gamma}
\ee
It is then straight-forward to evaluate the spatial Fourier-transform
\be \label{spatialF}
S^\pm_m(k) =  \frac{1}{L}\int_0^L dx \int_0^L dy \,e^{i (\pi-k) (x-y)}S^\pm_m(x,y), 
\ee
to obtain the
spectral weights $S_L$ in Eq.~(\ref{SL})
\be
S_L(\omega_m,k) = \frac{A_z L}{2 v} \left(S^+_m(k)-S^-_m(k)\right).
\label{SL2}
\ee  
In the case of odd $L$ the integrands $S_m^\pm(x,y)$ acquire an additional factor of 
$\cos \pi (x\pm y)/L$ from zero modes which reflects the parity symmetries of the wavefunctions.
Note that the spatial Fourier transform in Eq.~(\ref{spatialF}) 
dominates for antiferromagnetic wavevectors $k\approx \pi$, i.e.~small $q= k-\pi$. 
The expression for $S^\pm_m(x,y)$ from Eq.~(\ref{recursion}) contains products of
different $\gamma_\ell$ with the starting value $S_0^\pm$, so the spatial integral in 
Eq.~(\ref{spatialF}) can be evaluated exactly using known integrals as shown in the appendix.

In the following we use this procedure to efficiently calculate 
spectral weights $S_L(\omega_m,k)$ for comparison to numerics, for impurity averaging, and
for extracting the asymptotic behavior for long chains. 
However, it has been shown before that bosonization results can strongly depend on
non-linear corrections
\cite{imambekov2009,glazman2011,pereira2006,pereira2007,
pereira2008,cheianov2008,karrasch2015,eliens2016,karimi2011} or the cut-off procedure
\cite{meden1999,markhof2016},
so we first critically examine
if this approach gives correct results.
To this end we use DMRG \cite{white}
to calculate spectral weights in finite systems.  Using the  multi-targeting 
algorithm for spectral weights \cite{schneider2008} we can calculate the first 97 excitations,
which captures all nearly-degenerate multiplets up to the energy level $m=9$. Using $M=600$ DMRG states
gives an accuracy in the wavefunction of order $10^{-2}$ relative to exact results
from the $xx$-model.

A direct comparison between bosonization $S_L$ and numerics  $S_{\rm DMRG}$ is shown
in Fig.~\ref{fig:level8and9} for energy levels 
$m=8$ and $m=9$ in a finite system of $L=100$ with $K=0.8$.  
Without any fit the
agreement is surprisingly accurate and even captures details
like an alternating signal at $k=\pi$ with even and odd $m$ due to parity symmetry, 
which leads to overall oscillations. 
Due to the zero-mode prefactor the same alternation is
observed between even and odd lengths $L$. 
For a quantitative analysis we also compare the small 
error between DMRG $S_{\rm DMRG}$ and bosonization $S_L$
with the finite size correction relative to the
bulk behavior $S_\infty$ in Eq.~(\ref{bulk}) 
 by defining 
\be \Delta S \equiv \frac{S_L-S_{\rm DMRG}}{S_L-S_\infty}.\label{Delta}
\ee
Both the numerator and the denominator go to zero as $L\to \infty$, but the error to the numerics
vanishes quicker with $1/L$ as
shown in the lower right panel of Fig.~\ref{fig:level8and9} for $m=4$, $K=0.8$ and 
selected $k-$values, for which the denominator tends to be small. 

We are now in the position to efficiently calculate $S_L$ for a large range of 
$L$, $k$, and $\omega$ 
to average the signal in a randomly 
doped system.  An impurity density $p$ of non-magnetic sites gives
a distribution of chain lengths \cite{wessel,eggert2002}
$P(L) = p^2 (1-p)^L$ 
normalized so that $\sum L P(L) = 1-p$. 
The averaged signal $\bar S_p$ for typical experimental doping values
in Fig.~\ref{fig:level8and9}
shows that the signal at the divergence is strongly reduced relative to the undoped 
case $L\to \infty$
while significant spectral weight is observed just outside the dispersion $|vq| > \omega$.

\begin{figure}
\includegraphics[width=1\columnwidth,keepaspectratio]{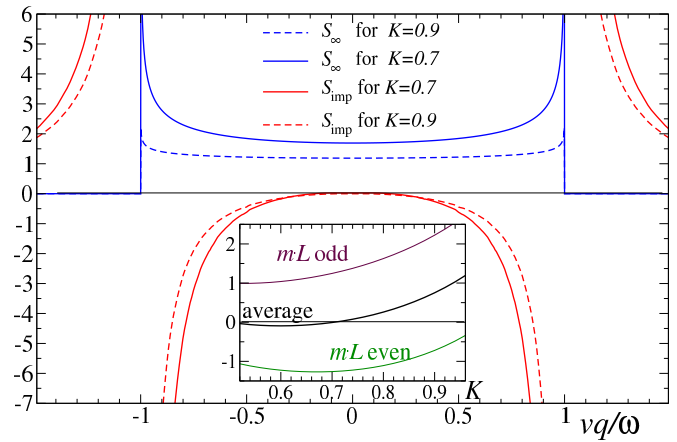}  
\caption{The rescaled signal $(\frac{\omega}{v})^{2-2K} S_{\rm \infty}$ and  $(\frac{\omega}{v})^{3-2K} S_{\rm imp}$
as functions of the scaling variable $vq/\omega$ for $K=0.7$
and $K=0.9$. {\it Inset:} 
Relative impurity contribution
$\omega \widehat S_{\rm imp}(\omega)/v \widehat S_\infty(\omega)$ from the
$k-$integrated signal as a function of $K$. } \label{results}
\end{figure}

It must be emphasized that the finite-size bosonization is completely divergence free.
For any finite or impurity-doped system we obtain a 
well-behaved finite signal even at $|vq| = \omega$,
so it is unclear in what situation the 
power-law divergence in Eq.~(\ref{bulk}) becomes relevant.  To answer this question we 
analyze the impurity correction $S_{\rm imp}$ relative to the
thermodynamic limit, which is defined as the first order 
in a $1/L$ expansion \cite{eggert1995}
\be \label{imp}
S_L(\omega,k) = S_\infty(\omega,k) + 
{L^{-1}} S_{\rm imp}(\omega, k) +{\cal O}\left(L^{-2}\right).
\ee
Based on the efficient calculation of spectral weights from 
Eqs.~(\ref{recursion})-(\ref{SL2}) we
can make a comprehensive finite-size scaling to determine $S_\infty$ and $S_{\rm imp}$ 
for different 
$\omega$ and $k$. 
Due to the scale-invariance of the underlying bosonization the resulting 
contributions in Eq.~(\ref{imp}) show perfect data collapse, so that $\omega^{2-2K}S_\infty$
and $\omega^{3-2K}S_{\rm imp}$ are only functions of the scaling variable $vq/\omega$
as shown in Fig.~\ref{results} for $K=0.7$ and $K=0.9$.  While $S_\infty$ 
is given by Eq.~(\ref{bulk}), we find that $S_{\rm imp} \propto \omega^{2K-3}$ increases even faster 
with decreasing $\omega$.  This is reminiscent of quantum wires, which also 
show boundary dominated spectral functions at low energies \cite{boundaryDOS}.
 Even 
more interesting is the strong divergence of the impurity part 
in Fig.~\ref{results}, which goes as
$\left||vq|-\omega\right|^{K-2}$
and implies a breakdown of the 
expansion in Eq.~(\ref{imp}) as $|vq| \to \omega$. 
In particular, summing over 
higher order corrections
in $1/L$ in Eq.~(\ref{imp}) would be required as $|vq|\to \omega$
with more and more divergent powerlaws, 
even though the final result must be 
{\rm finite} at the corresponding length as shown above.

Nonetheless, the expansion in Eq.~(\ref{imp}) is useful away from the divergences
in order to estimate the length-averaged signal to lowest order in $p$
\be \label{avg}
\bar S_p(\omega,k)
  \approx  E_1\!\left(\frac{p\pi v}{\omega}\right)\! S_\infty(\omega,k) + p
E_2\!\left(\frac{p\pi v}{\omega}\right)\! S_{\rm imp}(\omega,k), 
\ee
in terms of the Einstein functions
$E_1$ 
and $E_2$ 
\be E_1(x) = \frac{x^2 e^x}{(e^x-1)^2} \  \ {\rm and}\ \ E_2(x) = \frac{x}{e^x-1},
\label{weightfunctions}
\ee
that are derived 
in the appendix.
Both $E_1$ and $E_2$ become exponentially small for energies below
the average-length gap $\omega \ll \pi v/\bar L \equiv p \pi v$.
The suppression of bulk spectral weight with $E_1$ due to the finite size gaps 
was discussed and 
observed experimentally \cite{average,average2}, but we find that 
the additional redistribution of spectral weight becomes very important, which can be 
traced to the effect of boundary correlations.
The rescaled average $\omega^{2-2K} \bar S_p$ from Eq.~(\ref{avg}) is now a function of 
two scaling variables $vq/\omega$ and $ vp/\omega$. The corresponding data collapse holds
approximately also for the averages over 
all lengths shown in Fig.~\ref{fig:level8and9} above, 
so that the signal for a given $\omega$ can easily be generalized to other energies.

The averaged signal in  Fig.~\ref{fig:level8and9} and the impurity
correction in Fig.~\ref{results} show that the signal is strongly reduced 
for $|vq|<\omega$, while spectral weight is created for $|vq|>\omega$.
This invites the question if the $k-$integrated signal $\widehat S(\omega)$
at a given energy 
is overall increased or decreased or
even unchanged due to the boundaries.  This is relevant for 
neutron scattering experiments, which recently observed significant
changes of the spectral weight around $k\approx \pi$ depending on the doping \cite{neutron,neutron2}.
To calculate the integrated antiferromagnetic spectral weight 
$\widehat S$ as a function of $L$, we 
use the fact that
an integration over $k$ of Eq.~(\ref{spatialF}) leads to delta-functions $2\pi\delta(x-y)$, 
so it is possible to apply 
the recurrence relation in Eq.~(\ref{recursion}) for $S^\pm_m(x,x)$, which is
inserted into the corresponding spatial integral.  Finite size scaling gives
a bulk part $\widehat S_\infty(\omega) \propto \omega^{2K-1} $ which now decreases 
with decreasing $\omega$ corresponding to the integral of Eq.~(\ref{bulk}).  However, the 
impurity part $\widehat S_{\rm imp}(\omega) \propto \omega^{2K-2}$ 
increases with decreasing $\omega$, so we define the energy independent
ratio $\omega \widehat S_{\rm imp}(\omega)/v \widehat S_\infty(\omega)$, 
which is only dependent on $K$ (i.e.~$\Delta$)
as shown in the inset of Fig.~\ref{results}. 
Note that 
due to the alternation with $m$ and $L$ the impurity part is different
if $m\cdot L$ is even or odd, but the experimentally relevant average 
gives a finite and relatively small value.
Therefore, the corresponding
expansion and averaging in Eqs.~(\ref{imp}) and (\ref{avg}) work well to 
calculate the doping and energy dependence using the $k$-integrated data in Fig.~\ref{results} (inset).
The impurity part 
becomes negative at $K\alt 0.7$ i.e.~larger $\Delta$,
which may in part explain an additional depletion of spectral weight at low
energies, but the experimentally observed changes with different 
impurity types \cite{neutron} require more refined models beyond simple chain breaks.

Last but not least it is instructive 
to consider 
finite systems with periodic boundary conditions.  The starting values in Eq.~(\ref{start})
are now independent of position $c=\left(\frac{2\pi}{L}\right)^K$, so
 all integrals 
can be done directly.  As shown in the appendix  the recurrence relation leads to an analytical result for
all energies, lengths, and momenta 
\be \label{periodic}
S_L(\omega_m, k_l)
 = \frac{  A_z  L^2  \, c^2}{4 v \Gamma^2(K)}
\frac{\Gamma(\frac{m+l}{2}+K)}{\Gamma(\frac{m+l}{2}+1)}
\frac{\Gamma(\frac{m-l}{2}+K)}{\Gamma(\frac{m-l}{2}+1)}
\ee
where now $\omega_m=2\pi v m/L$ and also $k_l-\pi = 2\pi l/L$ is quantized 
due to periodicity with the condition that
$l$ and $m$ are either both even or both odd integers
and $|l|\leq m$. Therefore, there is no spectral 
weight for $v|k_l-\pi|>\omega_m$ in strong contrast to open boundary condition 
discussed above.  It is straight-forward to expand Eq.~(\ref{periodic})
in $1/L$ using Stirlings formula to obtain 
$S_\infty$ in Eq.~(\ref{bulk}) and a negative impurity part $S_{\rm imp}$.

In summary we have analyzed the structure factor of doped spin chain systems.  
Using bosonization and numerical DMRG, we see that doping leads to 
a significant shift of spectral weight from lower momenta to 
regions $v|k-\pi|> \omega$ in neutron scattering experiments, which would not show any signal
for infinite or periodic systems. 
The relative change from doping near the dispersion 
$|vq| \to \omega$ is
infinitely large, so that the first order impurity contribution 
diverges near the dispersion $|vq| \to \omega$ with a stronger powerlaw than the bulk
and 
a $1/L$ expansion from the thermodynamic limit 
always breaks down. 
Previous studies also 
found that the divergence in the thermodynamic limit is not universal, but instead
strongly dependent on either the cut-off procedure \cite{meden1999,markhof2016} or higher order terms 
and non-linear effects \cite{imambekov2009,glazman2011,pereira2006,pereira2007,
pereira2008,cheianov2008,karrasch2015,eliens2016,karimi2011}.
Naively, it could have been expected 
that bosonization works particularly well in the thermodynamic limit, but instead
it turns out that the finite-size theory
is much better controlled and quantitatively accurate even for $|vq| \to \omega$ as 
shown in Fig.~\ref{fig:level8and9}.  
From a technical point of view, the mode expansion for finite systems 
leads to finite sums, which can be efficiently evaluated using a recurrence relation
without the need for contour integral, asymptotic limits, non-linearities, 
or cut-off procedures.

It is fair to say that in one dimension it is always important to consider 
boundaries, since  
physical systems only contain finite chains 
even in the absence of doping \cite{takigawa}.  This is especially also true for
artificially created spin chains using surface structures \cite{sts1,sts2}, 
ion-traps \cite{ion}, or 
ultra-cold gases \cite{gases1,gases0,gases3}
as quantum simulators, where
measurements of energy and space resolved correlations 
are in principle possible \cite{gases2}. 

Finally we would also like to discuss the limitations and open questions which remain.
It is known that for higher energies 
than considered in Fig.~\ref{fig:level8and9}
higher order operators play a role, 
which lead to systematic corrections \cite{log1,eggert1992,log,sirker-bortz}.  
In particular, in the limit $\Delta\to 1$ it is well known that logarithmic corrections
lead to strong quantitative changes \cite{log1}.  Those log-corrections have not yet 
been fully understood for open 
boundary systems \cite{log,sirker-bortz} and are beyond the scope of this paper.  Nonetheless, 
preliminary DMRG simulations at $\Delta =1$ show that the strong 
transfer of spectral weight to $v|k-\pi|> \omega$ is a robust feature.

\begin{widetext}
\section{APPENDIX}


Here we review the bosonization and calculation of 
correlation functions for finite spin-1/2 $xxz$-chains
\begin{eqnarray} \label{H}
 H=J \sum_{i}^{} \left( S_i^xS_{i+1}^x +  
S_i^y S_{i+1}^y  + \Delta  S_i^z S_{i+1}^z \right)
\end{eqnarray}
with $L$ sites and open or periodic boundary conditions.
Using the correlation functions we want to 
calculate the dynamic structure factor
at low frequencies and 
near the antiferromagnetic wave-vector  $k\approx\pi$, which is given by
\begin{eqnarray}
S(\omega,k)& =& \frac{1}{L}\sum_{j,j^\prime} e^{-i k (j-j^\prime)} \int_{-\infty}^\infty dt \,e^{i \omega t}\langle S^z_j(t)S^z_{j^\prime}(0)\rangle \nonumber\\
& =& \Delta \omega \sum_{m\neq 0} S_L(\omega_m,k) \delta(\omega-\omega_m)
\label{lehmann2}
\end{eqnarray}
where in the last line we have used the Lehmann representation using
individual spectral weights
\be 
S_L(\omega_m, k) =
\frac{2 \pi}{\Delta \omega} |\langle \omega_m | S_k^z|0 \rangle |^2
\ee
at discrete energies $\omega_m= m \Delta \omega$.

\section{Bosonization and correlations for open boundary conditions}

The low-energy theory for the model in Eq.~(\ref{H})
is well described in the continuum limit by bosonic fields, which are rescaled
by  the square-root of the Luttinger parameter  $K=\pi/2(\pi-\theta)$ 
where $\cos\theta=\Delta$ \cite{review}.  
The free
Hamiltonian is given by 
\begin{eqnarray}
H=\frac{v}{2}\int_0^L dx[ \Pi(x)^2+(\partial_x \phi(x))^2].
\end{eqnarray}
where 
$v=J \pi \sin\theta/2 \theta$ is the spinon velocity and 
$\Pi$ is the momentum density conjugate to $\phi$, $[\phi(x),\Pi(y)]=i \,\delta(x-y)$.
Higher order corrections are well understood \cite{log1,sirker-bortz,bortz,eggert1992}, but are irrelevant for
low energies and long chains.

We are interested in the local $S^z$-operators, which can be expressed in terms of the bosons
\begin{eqnarray}\label{eq:Sz}
S^z(x,t)=\sqrt{\frac{K}{\pi}}\partial_x\phi(x,t)+A(-1)^{x}\sin\left(\sqrt{4\pi K} \phi(x,t)\right), 
\label{Sz}
\end{eqnarray}
where $A^2 = A_z/2$ is related to the amplitude of the asymptotic correlation functions, that  
is known from exact methods \cite{lukyanov2002}.

Open boundaries lead to the following 
mode expansion of the bosonic fields \cite{eggert1992,eggert2002} 
\begin{eqnarray}
\phi(x,t)=\hat{Q}\frac{2 \,x}{L} +\phi_{\rm osc}(x,t) 
\end{eqnarray}
with 
\begin{eqnarray}
\phi_{\rm osc}(x,t)=\sum_{\ell=1}^\infty \frac{1}{\sqrt{\pi \ell}} \sin{ \frac{\pi \ell x}{L}} 
\left(e^{-i  \frac{\pi \ell v t}{L}}\, b_{\ell} + e^{i  \frac{\pi \ell v t}{L}}\, b^{\dagger}_{\ell}\right).
\end{eqnarray}
The zero mode $\hat{Q}$ is given in terms of the total magnetization 
\be
S^z = \int_0^L \sqrt{\frac{K}{\pi}} \partial_x \phi = 
2 \sqrt{\frac{K}{\pi}}  \hat{Q}.
\ee
Note that those expressions agree with previous works \cite{review,eggert1992,eggert2002}, up to an overall phase shift 
$\phi_0$ in the boson, which is of no consequence.

For the dynamical structure factor near $k\approx \pi$ we are interested in the 
alternating part of the $S^zS^z$-correlation function
\bea
 \langle \sin (\sqrt{4\pi K} \phi(x,t)) \sin(\sqrt{4\pi K} \phi(y,0)\rangle
&  = &  \frac{1}{2}\left(G^+(x,y,t) - G^-(x,y,t)\right)
\eea
with 
\bea
G^\pm(x,y,t) & =&  \langle e^{i 2 \pi S^z (x\mp y)/L}\rangle
 \langle e^{i\sqrt{4 \pi K}\phi_{\rm osc}(x,t)} e^{\mp\sqrt{4 \pi K}\phi_{\rm osc}(y,0) }\rangle.
\label{G0}
\eea
The first factor gives different results for even chains $S^z=0$
and for odd chains $S^z=\pm 1/2$ \cite{eggert2002}
\be
\langle e^{i 2 \pi S^z (x\pm y)/L}\rangle = 
\left\{\begin{array}{lcl}
 1, & \phantom{nn}& L\text{  even} \\
 & & \\
\cos(\frac{\pi (x\pm y)}{L}), & \phantom{nn}& L\text{  odd} \\
\end{array} \right. 
\label{zerom}\ee
which reflects the different parity symmetry of the 
wavefunctions in even and odd chains.
For the second factor in Eq.~(\ref{G0}) it is useful to apply normal ordering
\be
\label{opnorm}
\exp\left({i\sqrt{4 \pi K}\phi_{\rm osc}(x,t)}\right)=
c(x) \exp\left({ i\sum_{\ell} e^{i \omega_l t} \frac{A_\ell^\dagger(x)}{ \sqrt{\ell}}}\right)
\exp\left({ i\sum_{\ell} e^{-i  \omega_\ell t} \frac{A_\ell(x)}{ \sqrt{\ell}}}\right)
\ee
where $\omega_\ell = \ell \Delta \omega$ with $\Delta \omega = \frac{\pi v}{L}$ and 
operators \cite{schneider2010}
\be 
A_\ell(x) = 2  \sqrt{K} \sin \frac{\pi \ell x}{L}\ b_\ell.
\ee
The prefactor is given via the Baker-Campbell-Hausdorff formula by
\be 
c(x) = \exp\left({-\sum_\ell \frac{2K}{\ell} \sin^2\frac{\pi \ell x}{L}}\right) ,
\ee
which is divergent.  However, using 
\be \sum_{\ell=1}^\infty q^\ell/\ell= -\log (1-q) \label{log} \ee
it is possible to capture the dependence on $L$ and $x$ correctly, so that
only an overall factor is dependent on the regularization, which we choose to be finite
by setting
\be
c(x) = \left(\frac{2L}{\pi}\sin{\frac{\pi  x}{L}}\right)^{-K}.
\label{c}
\ee
Therefore, upon using Baker-Campbell-Hausdorff again, the correlation 
functions in Eq.~(\ref{G0}) becomes
\begin{eqnarray}\label{G1}
G^\pm(x,y,t)=c(x)c(y)\exp\left(\sum_{\ell=1}\frac{\pm1}{\ell}e^{-i \,\omega_\ell t} \gamma_\ell(x,y)\right)
\end{eqnarray}
where we introduced the commutator
\be
\gamma_\ell(x,y)
= [A_\ell^{\phantom{\dagger}}(x),A_\ell^\dagger(y)] =
4 K \sin\frac{\ell \pi  x}{L} \sin\frac{\ell \pi  y}{L}.
\label{gamma2}
\ee
For odd chains, the additional factor in Eq.~(\ref{zerom}) must also be inserted.

At this point all information for the asymptotic behavior of the correlation 
function is known, which in fact can be expressed in closed form using Eq.~(\ref{log}) \cite{eggert1992,eggert2002,eggert1995,mattsson}
\be
G^\pm(x,y,t)
= c(x)c(y) \left[\frac{\sin \frac{\pi(x+y-v t)}{2L}\sin \frac{\pi(x+y+v t)}{2L}}{\sin \frac{\pi(x-y-v t)}{2L} \sin\frac{\pi(x-y+v t)}{2L}}\right]^{\pm K}
\label{G2}
\ee
for even $L$ (and by including the factor in Eq.~(\ref{zerom}) for odd $L$).
Note that we have normalized the correlation function so that
\begin{eqnarray}\label{normope}
G^+(x,y,t) \to \left((x-y)^2-\upsilon^2 t^2\right)^{-K} 
 \end{eqnarray}
in the thermodynamic limit away from the boundary.
The overall prefactor must be determined from exact methods \cite{lukyanov2002}, so that 
the normalization in Eqs.~(\ref{c}) and (\ref{normope}) is simply a matter of convenience.

\section{Fourier transform and recursive formula}
To calculate the dynamical structure factor it is useful to go back 
to Eq.~(\ref{G1}) in order to obtain the Fourier transformation in time. 
In accordance with the periodicity in $t$ this yields an expansion in delta functions
\begin{eqnarray}
\int_{-\infty}^\infty dt \, e^{i \omega t}\, G^\pm(x,y,t) = 2 \pi \sum_{m}  S_{m}^\pm (x,y) \delta(\omega- \omega_m ). 
\end{eqnarray}
where the discrete spectral weight for $\omega_m=m \Delta\omega = m \frac{\pi v}{L}$
 is determined by the functions $\gamma_l$ 
in a recursive way  \cite{schneider2010},
\begin{eqnarray}\label{recursion2}
S_{m}^\pm(x,y)=\frac{\pm 1}{m}\sum_{\ell=1}^m S_{m-\ell}^\pm(x,y) \,\gamma_\ell(x,y).
\end{eqnarray}
which simply follows from partial integration.
This equation defines the recursion formula, which allows to calculate any individual 
spectral weight as a sum of the previous ones from starting values
$S_{0}^\pm(x,y)=c(x)c(y)$ (and including Eq.~(\ref{zerom}) for odd $L$).  
Note that this is much easier than an integration over Eq.~(\ref{G2}) which would require a 
small imaginary cutoff for the time and a complicated
contour integration.

For the spatial Fourier transform we define
\begin{eqnarray}\label{spatialF2}
S^\pm_m(k)=\frac{1}{L}\int_0^L dx \int_0^L dy \,e^{i (\pi-k) (x-y)}S^\pm_m(x,y)
\end{eqnarray}
where the
shift of the wavevector by $\pi$ follows from the alternating factor 
in Eq.~(\ref{Sz}).
Using $S_m(k)=\frac{A_z}{4}\left(S^+_m(k)-S^-_m(k)\right)$
we obtain
\begin{eqnarray}
S(\omega,k)=2\pi \sum_m S_{m}(k) \,\delta(\omega- \omega_m ).
\end{eqnarray}
Since the integrand $S^\pm_m(x,y)$ in Eq.~(\ref{spatialF2}) only involves a sum of 
exponentials $\exp(i \ell_x \pi x/L)$ and $\exp(i \ell_y \pi y/L)$ according 
to Eqs.~(\ref{gamma2}) and (\ref{recursion2}), it is possible to perform the integral for
each such term 
analytically together with the prefactor $c(x)$ in Eq.~(\ref{c}) by using
\begin{align}
\int_0^L dx &\frac{e^{i\frac{\pi}{L}q x}}{\left(\sin \frac{\pi x}{L}  \right)^{K}} =
 \frac{\pi  e^{ i \pi  q/2} 2^K L \csc (\pi  K)}{\Gamma (K) \Gamma \left(q/2-K/2+1)\right) \Gamma \left(-q/2-K/2+1\right)}
\label{int2}
\end{align} 
for $K<1$ and analogously for the integration over $y$. 
In the summation of Eq.~(\ref{recursion2}) we therefore keep track of the prefactors for 
each pair $(\ell_x, \ell_y)$ for each level $m$ 
and then add up the exactly known integrals as a function of $k$ in Eq.~(\ref{int2})
   in the end.

\section{Periodic boundary conditions}

The recursive approach is particularly simple for periodic boundary conditions. 
In this case the system is translationally invariant, so that $G^+(x,y,t)$  is a function of $x-y$ and $t$ only and $G^-(x,y,t)$ vanishes. The prefactor is constant  
$c(x)=c=\left(\frac{2\pi}{L}\right)^K$. 
It is then convenient to introduce light-cone coordinates $z=v t-(x-y)$ and $\bar{z}=v t +x -y$ such that the correlation function factorizes 
\begin{eqnarray}
e^{-i k(x-y)}e^{i\omega t} G^+(x,y,t)= e^{i u z} e^{i \bar{u} \bar{z}}G(z)G(\bar{z})
\end{eqnarray} where $k$ is measured relative to $\pi$ and 
  \begin{eqnarray}\label{Gfact}
G(z)=c\,\exp\left(\sum_{\ell}\frac{1}{\ell}e^{-i \,\frac{2\pi}{L} \ell z } \gamma\right) 
\end{eqnarray}
with $\gamma=K$. 
The double Fourier transform  in $z$ and $\bar{z}$ with frequencies 
$u=\frac{1}{2}(\frac{\omega}{v} +k )$ and $\bar{u}=\frac{1}{2}(\frac{\omega}{v} -k)$, respectively, can  then be performed directly by applying the recursion formula in Eq.~(\ref{recursion2}) to the contributions of right-movers and left-movers separately.  
Due to periodicity with $L$ in $z$ and $\bar z$, the values for both $u$ and $\bar u$ are
quantized
\be
u =  \frac{2\pi}{L} n\ \ \ \ \ \bar u =  \frac{2\pi}{L} \bar n
\ee 
The boundaries of the integrals transform as follows: 
\begin{eqnarray}
\frac{1}{L}\int_0^L dx \int_0^L  dy\to\frac{1}{L}\int_0^L dy \int_{-y}^{L-y}  dr=\int_{0}^{L}  dr
\end{eqnarray}
 for integrands independent of $y$ and $L$-periodic in $r$. Furthermore we use 
\begin{eqnarray}
\int_0^L dr \int_{-\infty}^{\infty}dt&\to&\frac{1}{2\upsilon}\int_{-\infty}^\infty dz\int_{-z}^{2L-z} d\bar{z}
= \frac{1}{2\upsilon} \int_{-\infty}^\infty dz \int_{0}^{2L} d\bar{z}
\end{eqnarray}
 for integrands invariant under $\bar{z}\to \bar{z}+L$.   
Since $\gamma=K$ is independent of $\ell$ in Eq.~(\ref{Gfact}), the 
recursion can be solved exactly to give a ratio of gamma functions \cite{schneider2010}, i.e.
 \begin{eqnarray}\label{RatioG}
\int_{-\infty}^\infty e^{i u z} G(z) dz=\frac{2 \pi \, c}{\Gamma(K)} \sum_{n}\frac{\Gamma(n+K)}{\Gamma(n+1)}\delta\left(u-\frac{2\pi}{L}n\right)
\end{eqnarray}
and 
\begin{eqnarray}
\int_{0}^{2L} e^{i \bar{u} \bar{z}} G(\bar{z}) d\bar{z}
&=&\frac{2 L  \, c}{\Gamma(K)} \sum_{\bar n}\frac{\Gamma(\bar n+K)}{\Gamma(\bar n+1)}
\delta_{\bar n, \bar{u}L/2\pi}  
\end{eqnarray}
for the integration over $\bar{z}$.
Now using the exact result for the asymptotic amplitude of the alternating correlation functions $A_z$ from Ref.~\cite{lukyanov2002} we obtain
\begin{eqnarray}
\!\!\!\!\!\!\!\!\!\!\!\!\!\!\!\!\!\!\!\!\! S(\omega,k)&=&\frac{ \pi A_z  L  \, c^2}{2v \Gamma^2(K)} \sum_{n,\bar n}\frac{\Gamma(n+K)}{\Gamma(n+1)}\frac{\Gamma(\bar n+K)}{\Gamma(\bar n+1)}
  \delta\left(u - \frac{2\pi}{L}n\right) \delta_{\bar u, 2 \pi \bar n/L}\\
&=&\frac{ \pi A_z  L  \, c^2}{2 \Gamma^2(K)} 
\sum_{m}\sum_{l=-m}^m\frac{\Gamma(\frac{m+l}{2}+K)}{\Gamma(\frac{m+l}{2}+1)}
\frac{\Gamma(\frac{m-l}{2}+K)}{\Gamma(\frac{m-l}{2}+1)}
   \delta\left(\omega - \frac{2\pi v m}{L}\right) \delta_{k, 2 \pi l/L}
\end{eqnarray}
where the sum over $l$ goes in steps of two, so that $l= n-\bar n$ and $m=n + \bar n$ 
are either both even or both odd and $|l|\leq m$.
Comparing with Eq.~(\ref{lehmann2}) we can write for quantized frequencies $\omega_m = m \Delta\omega = m\frac{2 \pi v}{L}$ and momenta $k_l-\pi=l \frac{ 2 \pi}{ L}$
\be
S_L(\omega_m, k_l) 
=\frac{  A_z  L^2  \, c^2}{4 v \Gamma^2(K)} 
\frac{\Gamma(\frac{m+l}{2}+K)}{\Gamma(\frac{m+l}{2}+1)}
\frac{\Gamma(\frac{m-l}{2}+K)}{\Gamma(\frac{m-l}{2}+1)}
\ee
Stirling's formula for large arguments $\Lambda$ gives
\be
\frac{\Gamma(\Lambda+K)}{\Gamma(\Lambda+1)} \approx \Lambda^{K-1} \left(1+\frac{K(K-1)}{2\Lambda} + {\cal O}\left(\frac{1}{\Lambda^2}\right)\right)
\ee
so that to leading order we find the bulk behavior in the thermodynamic limit
\begin{eqnarray}
S_\infty(\omega,q+\pi) = \frac{\pi^2 A_z}{2 v \Gamma^2 (K)}2^{2-2K} \left(\frac{\omega^2}{\upsilon^2}-q^2 \right)^{K-1}\ \ \ {\rm for} \ \ v|q|<\omega,
\label{eq:Smpbc}
\end{eqnarray} 
where we get a factor of 2 due to the fact that the quantization 
of $k_l$ jumps in steps of two at a given $m$. 
 The analogous analysis can be made for odd $L$ where the 
prefactor in Eq.~(\ref{zerom}) basically gives the sum of two contribution with 
the $k$-quantization changed by one $l\to l\pm 1$.

\section{Integrated spectral weight}

For the total spectral weight near the antiferromagnetic wave vector, we
can integrate the contribution from the alternating correlation function
\begin{eqnarray}
\widehat S(\omega)= \int dk S(\omega,k)
\end{eqnarray} 
where the integral is taken in the vicinity of $k=\pi$. 
Let us also define the integrated spectral weight at discrete energies by
\begin{eqnarray}
\widehat S(\omega)=2 \pi \sum_m \widehat S_m \delta(\omega-\omega_m).
\end{eqnarray}
with $\widehat S_m=\frac{A_z}{4}(\widehat S_m^+-\widehat S^-_m)$.
Integrating Eq.~(\ref{spatialF2}) over $k$ generates a delta function
 $2\pi \delta(x-y)$ such that one spatial integration can be trivially performed and $S_m$ simplifies to
\begin{eqnarray}\label{eq:Smint}
\widehat S_m=\frac{\pi A_z }{2  L} \int_0^L dx \left(S_{m}^+ (x,x)-S_{m}^-(x,x)\right).
\end{eqnarray}  
The functions $S_m^+(x,x)$  are generated recursively via Eq.~(\ref{recursion2}).
In case of periodic boundary conditions -- since $\gamma_l(x,x)=2K$ is independent of $l$ -- the recursion can again be solved exactly. From Eq.~(\ref{RatioG}) we find
\begin{eqnarray}
\widehat S_m= \frac{\pi A_z c^2}{2 \Gamma(2 K)} \frac{\Gamma(m+2 K)}{\Gamma(m+1)}.
\end{eqnarray}
The bulk power law for the $k$-integrated structure factor is
\begin{eqnarray}
\widehat S_\infty(\omega)=\frac{\pi^2 A_z}{v \Gamma(2K)}\,\left(\frac{\omega}{v}\right)^{2K-1}.
\end{eqnarray}
Note that this result can also be obtained by directly integrating Eq.~(\ref{eq:Smpbc}). 

\section{Averaging over chain lengths}

For a doping density of $p = N_{imp}/N$ missing sites, the probability of finding a linear 
segment of length $L$ is \cite{wessel}
\be 
P(L) = p^2 (1-p)^L \approx p^2 \exp(- L p),
\ee
which is normalized so $N \sum P(L) = N_{imp}$. The probability of a single site to belong 
to a segment of length $L$ is $L P(L)$ which is normalized so that
$N \sum L P(L) = N- N_{imp}$, which excludes the missing sites.  
In the limit of large chains or small doping,  the sums can be converted
 to integrals since the signal does not change
significantly as a function of length 
so that $\int dL\  P(L) = p$ and $\int dL \ L P(L) = 1$.

For a segment of length $L$ we use the Lehmann representation in Eq.~(\ref{lehmann})
in order to define the average signal
\be
\bar S(\omega,k) =
    \sum_L\,  P(L) L S(\omega,k) \\
 \approx  \int dL \, P(L) \sum_m \pi v S_L(\omega_m,k) \delta(\omega-\omega_m)
\ee
which allows us to average separately
over the bulk and impurity contributions in the $1/L$ expansion from the thermodynamic limit
\be S_L (\omega_m,k) \approx S_\infty(\omega_m,k) +  \frac{1}{L} S_{\rm corr}(\omega_m,k)
+{\cal O}\left(\frac{1}{L^2}\right).
\label{corr}
\ee
 For the bulk average we find
\bea 
\bar S_\infty & = & \int_0^\infty dL\  \pi v p^2  e^{-L p} \sum_m S_\infty(\omega_m) 
\delta\left(\omega - m\frac{\pi v}{L}\right) \\
& = & \sum_m \int_0^\infty d\nu \  p^2 \frac{m \pi^2 v^2}{\nu^2} e^{-p m \pi v/\nu} S_\infty(\omega) \delta\left(\omega - \nu\right)\\
& = & \sum_m  \frac{m p^2 \pi^2 v^2}{\omega^2} e^{-p m \pi v/\omega} S_\infty(\omega)\\
& = &E_1(\pi v p/\omega) S_\infty(\omega) .
\eea
upon using the substitution $L= \frac{\pi v m}{\nu}$ and $dL = - d\nu \, \frac{m \pi v}{\nu^2}$.  Here
\be
E_1(y) = \sum_m m y^2 e^{-m y} = \frac{y^2 e^y}{(e^y-1)^2} \ee
is the Einstein function of the scaling 
variable $y= p \pi v /\omega$ which measures the "average-length" gap $v\pi/\bar L$ compared to $\omega$ \cite{average,average2}.
For the average impurity correction  
we use the same substitution $L= \frac{\pi v m}{\nu}$ and $dL = - d\nu \, \frac{m \pi v}{\nu^2}$
\bea 
\bar S_{\rm imp} & = & \int_0^\infty dL\  \frac{\pi v p^2  e^{-L p}}{L} \sum_m S_{\rm imp}(\omega_m) \delta\left(\omega - m\frac{\pi v}{L}\right) \\
& = & \sum_m \int_0^\infty d\nu\  p^2 \frac{\pi v}{\nu} e^{-p m \pi v/\nu} S_{\rm imp}(\omega) \delta(\omega - \nu)\\
& = & \sum_m  \frac{ p^2 \pi v}{\omega} e^{-p m \pi v/\omega} S_{\rm imp}(\omega)\\
& = & p E_2(\pi v p/\omega) S_{\rm imp}(\omega). 
\eea
which is proportional to $p$ and the scaling function
\be
E_2(y) = \sum_m y e^{-m y} = \frac{y}{e^y-1}.
\ee

\end{widetext}

\begin{acknowledgments}
This work was supported by the Deutsche Forschungsgemeinschaft (DFG) via the research
centers SFB/TR49 and SFB/TR185 and by the Studienstiftung des deutschen Volkes.
\end{acknowledgments}

\end{document}